\documentclass[12pt,letterpaper]{article}
\usepackage{amsmath,amssymb}
\usepackage{graphicx,color}
\usepackage{verbatim}

\usepackage[bf]{caption}

\usepackage{epsfig}

\usepackage{amsmath,amssymb}
\usepackage{graphicx,color}
\usepackage{bbm}

\usepackage[breaklinks,bookmarks=true,hyperfigures=true]{hyperref}
\usepackage[nosort]{cite} %
\usepackage[bulletsep]{collref} 

\newcommand{\rr}{\mathbb{R}}

\newcommand{\be}{\begin{equation}}
\newcommand{\ee}{\end{equation}}
\newcommand{\ba}{\begin{aligned}}
\newcommand{\ea}{\end{aligned}}
\newcommand{\ben}{\begin{displaymath}}
\newcommand{\een}{\end{displaymath}}
\newcommand{\bea}{\begin{eqnarray}}
\newcommand{\eea}{\end{eqnarray}}
\newcommand{\bean}{\begin{eqnarray*}}
\newcommand{\eean}{\end{eqnarray*}}
\newcommand{\f}{\frac}
\newcommand{\p}{\partial}

\newcommand{\la}{\langle}
\newcommand{\ra}{\rangle}

\renewcommand{\l}{\lambda}
\renewcommand{\th}{\theta}
\renewcommand{\a}{\alpha}

\renewcommand{\b}{\beta}

\newcommand{\g}{\gamma}

\newcommand{\e}{\epsilon}
\newcommand{\s}{\sigma}

\newcommand{\m}{\mu}

\newcommand{\om}{\omega}

\definecolor{green}{rgb}{0,0.5,0}

\renewcommand{\p}{\partial}

\ifx\href\asklfhas\newcommand{\href}[2]{#2}\fi

\definecolor{pink}{rgb}{0.7,0,0.7}

\newcommand{\AdSxS}{\ensuremath{{\rm AdS} \times {\rm S}}}
\newcommand{\AdSfiveS}{\ensuremath{{\rm AdS}_5 \times {\rm S}^5}}
\newcommand{\AdSthreeS}{\ensuremath{{\rm AdS}_3 \times {\rm S}^3}}

\newcommand{\mI}{\mathbbm{1}}	

\newcommand{\fkT}{{\ensuremath{\mathfrak{t}}}}

\newcommand{\nn}{\nonumber}

\newcommand{\calR}{{\ensuremath{\mathcal{R}}}}

\newcommand{\ket}[1]{{\vert{#1}\ra}}

\newcommand{\PSU}{{\ensuremath{\text{PSU}}}}
\newcommand{\SU}{{\ensuremath{\text{SU}}}}

\newcommand{\psu}{{\ensuremath{\mathfrak{psu}}}}
\newcommand{\su}{{\ensuremath{\mathfrak{su}}}}

\long\def\symbolfootnote[#1]#2{\begingroup
\def\thefootnote{\fnsymbol{footnote}}\footnote[#1]{#2}\endgroup}

\begin{document}

\begin{titlepage}       
\hfill
\parbox[t]{4cm}{\texttt{HU-EP-14/39 \\ 
	NORDITA-2014-113}}

\vspace{1.6cm}

\begin{center}

{\Large \bf Isometry Group Orbit Quantization \\of Spinning Strings in AdS$_3\times$S$^3$}

\vspace{1.2cm}

\renewcommand{\thefootnote}{\fnsymbol{footnote}}
\textsc{Martin Heinze,$^{a}$\footnote[2]{{Visiting PhD student at Nordita, KTH Royal Institute of Technology and Stockholm University,
Roslagstullsbacken 23, 106 91 Stockholm, Sweden}} 
George Jorjadze$^{a,\,b}$, Luka Megrelidze$^c$
}
\renewcommand{\thefootnote}{\arabic{footnote}}
\\[.6cm]

{\it ${}^a$Institut f{\"u}r Physik, Humboldt-Universit{\"a}t zu Berlin,}\\
{\it IRIS-Adlershof, Zum Gro{\ss}en Windkanal 6, 12489 Berlin, Germany}\\[3mm]
{\it ${}^b$Free University of Tbilisi and Razmadze Mathematical Institute,}\\
{\it University Campus of Digomi,\\ Agmashenebeli Alley 240, 0159, Tbilisi, Georgia}\\[3mm]
{\it ${}^c$Ilia State University,}\\
{\it K. Cholokashvili Ave 3/5, 0162, Tbilisi, Georgia}\\[.6cm]
\{\texttt{martin.heinze, george.jorjadze}\}\texttt{@physik.hu-berlin.de}\\
\texttt{luka.megrelidze.1@iliauni.edu.ge}

\vspace{1.5cm}

\end{center}

\centerline{{\bf{Abstract}}}
\vspace*{5mm}
\noindent
	Describing the bosonic AdS$_3\times$S$^3$ particle and string in SU$(1,1)\times$SU$(2)$ group variables, we provide a Hamiltonian treatment of the isometry group orbits of solutions via analysis of the pre-symplectic form. For the particle we obtain a one-parameter family of orbits parameterized by creation-annihilation variables, which leads to the Holstein-Primakoff realization of the isometry group generators. The scheme is then applied to spinning string solutions characterized by one winding number in AdS$_3$ and two winding numbers in S$^3$. We find a two-parameter family of orbits, where quantization again provides the Holstein-Primakoff realization of the symmetry generators with an oscillator type energy spectrum. Analyzing the minimal energy at strong coupling we verify the spectrum of short strings at special values of winding numbers.

\vspace{15pt}

\end{titlepage}

\newpage

\subsection*{Introduction and Conclusion}

	Finding the energy spectrum of string excitations in $\AdSxS$ backgrounds is one of the major goals in the study of the AdS/CFT correspondence \cite{Maldacena:1997re, Witten:1998qj, Gubser:1998bc}. For the cases of $\AdSfiveS$ and ${\rm AdS}_4 \times \mathbb{CP}^3$ a solution of the spectral problem has been proposed recently in terms of the so-called quantum spectral curve \cite{Gromov:2013pga, Gromov:2014caa, Cavaglia:2014exa},
	which though heavily relies on the conjectured quantum integrability, for a review see \cite{Beisert:2010jr}. Calculation of the string spectrum by first principles appears intricate and since the pioneering works \cite{Berenstein:2002jq, Frolov:2002av, Gubser:2002tv} the main considerations were restricted to semiclassical analysis around solutions of string dynamics, see the reviews \cite{Arutyunov:2009ga, Tseytlin:2010jv, McLoughlin:2010jw}. In addition, these studies require that some of the $\psu(2,2|4)$ charges diverge in the 't Hooft coupling as $\sqrt{\l} \gg 1$, usually corresponding to the classical string solutions becoming long, whereas for finite charges, the short string regime, the analysis formally breaks down. 

	As observed in \cite{Passerini:2010xc}, this subtlety seems to be connected to the particular role played by the string zero-modes, which obtain a mass-term determined by the non-zero-mode excitations and which for short strings scale differently in $\l$ than the non-zero-modes. Therefore, working in the bosonic subsector and using static gauge \cite{Jorjadze:2012iy}, in \cite{Frolov:2013lva} a generalization of the pulsating string \cite{deVega:1994yz, Minahan:2002rc} was constructed, which explicitly allowed for unconstrained $\AdSfiveS$ zero-modes. This so-called single-mode solution showed classical integrability and invariance under the isometries even at the quantum level.  Heuristically taking into account supersymmetric corrections, indeed the energy of the lowest excited state dual to a member of the Konishi multiplet was recovered up to order $\l^{-1/4}$.

	The present work should certainly be seen in this context. Note that as the single-mode solution \cite{Frolov:2013lva} is invariant under the isometries it is nothing but the ${\rm SO}(2,4)\times {\rm SO}(6)$ group orbit of the pulsating string solution \cite{deVega:1994yz, Minahan:2002rc} constructed in ${\rm AdS}_3$. Therefore, to devise similar systems one can consider the isometry group orbits of different well known string solutions and to find the supersymmetric generalization one should construct the orbits of the full symmetry group, $\PSU(2,2|4)$ in the case of $\AdSfiveS$.

	The Kirillov-Kostant-Souriau theory of co-adjoint orbit quantization is a powerful tool, see for example the seminal work \cite{Alekseev:1988ce}, and there is yet another reason to be interested in this method. As the bosonic string zero-modes become massive the work \cite{Frolov:2013lva} benefited immensely from thorough understanding of the massive bosonic particle in $\AdSxS$ \cite{Dorn:2005ja, Dorn:2010wt}. Hence, for a generalization to the full superstring one should also expect that at least some of the fermionic zero-modes obtain a mass and that knowledge of the massive $\AdSxS$ superparticle will be advantageous. However, even for the massless case our understanding of this system seems unsatisfactory, where for the case of $\AdSfiveS$ \cite{Metsaev:1998it} progress has been made in \cite{Metsaev:1999kb, Metsaev:1999gz, Horigane:2009qb, Siegel:2010gm}.

	In the present paper we describe the dynamics of the bosonic $\AdSthreeS$ particle and string in $\SU(1,1)\times\SU(2)$ group variables. After fixing our notation, we construct the isometry group orbits of a point particle sitting in the center of ${\rm AdS}_3$ and rotating in ${\rm S}^3$ and devise Hamiltonian treatment by analyzing the corresponding pre-symplectic 1-form. We find a one-parameter family of orbits naturally parametrized in creation-annihilation variables, which yields a Holstein-Primakoff realization of the isometry algebra \cite{Holstein:1940zp, Dzhordzhadze:1994np} and results in an oscillator-type energy spectrum. Hence, with relative ease we acquire exact quantization of the $\AdSthreeS$ particle, which shows consistency with previous results \cite{Jorjadze:2012jk}. By this, it seems plausible that quantization of the bosonic particle in other $\AdSxS$ spaces could be achieved by similar means. More interestingly however, quantization of the $\AdSxS$ superparticle, massless or massive, 
should be feasible by investigation of the supergroup orbits.

	Next, in the spirit of the single-mode string \cite{Frolov:2013lva}, we apply the orbit method to the spinning string solutions introduced in \cite{Frolov:2003qc}, see also \cite{Frolov:2003tu, Arutyunov:2003za} for more details. Following essentially the same steps as for the particle, the isometry group orbits are characterized by two parameters and the winding numbers of the spinning string. Investigation of the pre-symplectic 1-form again prompts a description in creation-annihilation variables, giving an oscillator-type realization of the symmetry generators and a corresponding spectrum. This yields exact quantization of the system, where in comparison to the particle one has more freedom in the Casimir numbers.

	However, the exact formula for the minimal energy $E_0$ turns out to be rather involved. Moreover, we expect our findings to match the result for the full superstring at leading order in strong coupling, $\l \gg 1$, only. Therefore, we conclude by investigating the minimal energy in this limit. As a check of our method, we study the different possibilities for the winding numbers and consistently identify long and short string solutions, characterized by their typical scaling behavior in 't Hooft coupling, $E_0\propto\l^{1/2}$ and $E_0\propto\l^{1/4}$, respectively.

	The main goal of this work is to demonstrate the applicability of the quantization scheme utilizing the isometry group co-adjoint orbits for a well known problem of current interest, namely quantization of particles and classical string solutions in $\AdSxS$ spaces. We are looking forward to extend our analysis by exploring the orbit method for supergroups, which hopefully give new insights on the spectral problem, especially in the limit of short strings.

	In particular, the ${\rm AdS}_5/{\rm CFT}_4$ duality has recently sparked extensive studies in related models. The machinery developed for $\AdSfiveS$ is currently adapted to less supersymmetric spaces \cite{Zarembo:2010sg, Wulff:2014kja}, viz., the superstring theory in ${\rm AdS}_2 \times {\rm S}^2 \times {\rm T}^6$ \cite{Zhou:1999sm, Berkovits:1999zq} and $\AdSthreeS \times {\rm M_4}$ \cite{Brown:1986nw, Seiberg:1999xz, Boonstra:1998yu, Maldacena:1997re}. Especially, in \cite{Hernandez:2014eta} the $\AdSthreeS$ spinning string studied in this work has been investigated in the presence of a NS-NS flux.
	
	Another prevailing topic is the investigation of the $q$-deformed $\AdSfiveS$ superstring, which was first discovered by tracing its integrability structure \cite{Beisert:2008tw, Beisert:2011wq}. Only recently the corresponding space-time has been understood \cite{Delduc:2013fga, Delduc:2013qra, Arutyunov:2013ega} and the generalization to other $\AdSxS$ spaces has been discussed in \cite{Hoare:2014pna}. However, the dual field theory is still unknown and instead of a conformal boundary the space-time shows a singularity, which seems to repell long string solutions \cite{SFatIGST2014, Kameyama:2014vma}.

	
	We are eager to see, whether the presented method shows to be useful in both of these contexts.

\subsection*{Notation and Conventions}

	Let us denote  coordinates of $\rr^{2,2}$ and $\rr^4$ by $(X^{0'},X^0,X^1,X^2)$ and $(Y^1,Y^2,Y^3,Y^4),$ respectively. The ${\rm AdS}_3$ and ${\rm S}^3$ spaces are defined by the embedding conditions
	\be\ba\label{AdS-S conditions}
		&X\cdot X = (X^1)^2+(X^2)^2-(X^{0'})^2-(X^0)^2=-1~,  \\
		&Y\cdot Y = (Y^1)^2+(Y^2)^2+(Y^3)^2+(Y^4)^2=1~.
	\ea\ee
	One identifies ${\rm AdS}_3$ with ${\rm SU}(1,1)$ and ${\rm S}^3$ with ${\rm SU}(2)$ by defining the group elements
	\be\label{g=Y}
		g = \begin{pmatrix}
			\,X^{0'}+iX^0 &X^1-iX^2\, \\ \,X^1+iX^2& X^{0'}-iX^0\,
		\end{pmatrix}, \qquad\quad 
		\tilde g = \begin{pmatrix}
			Y^4+iY^3 &Y^2+iY^1\, \\-Y^2+iY^1&Y^4-iY^3\,
		\end{pmatrix},
	\ee
	where generally, due to their similarity, quantities corresponding to ${\rm SU}(2)$ are denoted as the ones of ${\rm SU}(1,1)$, just with tildes.

	We use the following basis of the $\mathfrak{su}(1,1)$ algebra
	\begin{equation}\label{su(1,1) basis}
		\fkT_{0}=i{ \boldsymbol\s}_{3}~, \qquad\quad \fkT_{1}={\boldsymbol\s}_{1}~, \qquad\quad
		\fkT_{2}={\boldsymbol\s}_{2}~,
	\end{equation}
	with $\{\boldsymbol{\s}_1, \boldsymbol{\s}_2, \boldsymbol{\s}_3\}$ the Pauli matrices, such that the generators $\fkT_a$ satisfy the relations
	\begin{equation}\label{tt=}
		\fkT_a \,\fkT_b =\eta_{a b}\,\mI-\epsilon_{a b}\,^{c}\,
		\fkT_{c}~, \qquad\qquad\text{for}\quad a,b,c=0,1,2~.
	\end{equation}
	Here $\mI$ is the unit matrix, $\eta_{a b}=\text{diag}(-1,1,1)$ and
	$\epsilon_{a b c}$ is the Levi-Civita tensor, with
	$\epsilon_{012}=1$. The inner product defined by
	$\la\, \fkT_a\,\fkT_b\,\ra\equiv\frac{1}{2}\,
	\mbox{tr}(\fkT_a\,\fkT_b)
	=\eta_{a b}$
	provides the isometry between $\mathfrak{su}(1,1)$ and 3d Minkowski space,
	since for ${\mathfrak{u}}=u^a\,\fkT_a$
	one gets $\la\, {\mathfrak{u}}\,{\mathfrak{u}}\,\ra = u^a\,u_a$.
	Then, ${\mathfrak{u}}$ can be timelike, spacelike or lightlike as the corresponding 3d vector $(u^{_0},u^{_1},u^{_2}).$

	A standard basis in $\su(2)$ is given by  ${\tilde\fkT}_j = i\boldsymbol{\sigma}_j\,$ and one has
	\begin{equation}\label{ss=}
		{\tilde\fkT}_i\,{\tilde\fkT}_j=-\delta_{i j}\,{\bf I}-\epsilon_{i j k}\,{\tilde\fkT}_k~, \qquad\qquad\text{for}\quad i,j,k=1,2,3~.
	\end{equation}
	Hence,  $\su(2)$, with inner product $\la{\tilde\fkT}_i\,{\tilde\fkT}_j\ra
	= -\frac{1}{2}\,\mbox{tr}({\tilde\fkT}_i\,{\tilde\fkT}_j)=\delta_{ij}$, is isometric to $\rr^3$, i.e.,
	$\la\, \tilde{\mathfrak{u}}\,\tilde{\mathfrak{u}}\,\ra=\tilde u_j\,\tilde u_j$, where $\tilde u_j=\la\, {\tilde\fkT}_j\,\tilde{\mathfrak u}\,\ra$.

	The matrices $g$ and $\tilde g$ in \eqref{g=Y} and their inverse group elements can be written as
	\be\ba\label{decomposition}
		&g=X^{0'}\,\mI +X^a \,\fkT_a ~,
		\qquad ~~~\tilde g=Y^4\,\mI +Y^j\,{\tilde\fkT}_j~,\\
		&g^{_{-1}}=X^{0'}\,\mI -X^a \,\fkT_a ~, \qquad \tilde g^{_{-1}}=Y^4\,\mI -Y^j\,{\tilde\fkT}_j~,
	\ea\ee
	and from \eqref{tt=} and \eqref{ss=} one obtains the following relations between the length elements
	\be\label{dg=dY,dh=dX}
		\la\, g^{_{-1}}\text{d} g\,\,g^{_{-1}}\text{d}g\,\ra=\text{d}X\cdot \text{d}X~,~~~~~~
		\la\,\tilde g^{_{-1}}\text{d}\tilde g\,\,\tilde g^{_{-1}}\text{d} \tilde g\,\ra=\text{d} Y\cdot \text{d} Y~.
	\ee
	The isometry transformations are therefore given by the left-right multiplications
	\be\label{isometry tr}
		g\mapsto g_{l}\,g\,g_{r}~,~\quad \quad \tilde g\mapsto \tilde g_{l}\,\tilde g\,\tilde g_{r}~.
	\ee

\subsection*{The Particle in ${\rm SU}(1,1)\times{\rm SU}(2)$}
	The dynamics of a particle in ${\rm SU}(1,1)\times {\rm SU}(2)$ is described by the action
	\be\label{particle action 1}
		S=\int \text{d}\tau\,\left(\frac{1}{2\xi}\Big(
			\la\, g^{_{-1}}\dot{g}\,g^{_{-1}}\dot{g}\,\ra +
			\la\, \tilde g^{_{-1}}\dot{\tilde g}\,\tilde g^{_{-1}}\dot{\tilde g}\,\ra\Big)
		-\frac{\xi \m_{0}^2}{2}\right)~,
	\ee
	where $\xi$ plays the role of the world-line einbein
	and $\m_{0}$ is the particle mass. In the first order formalism, this action is equivalent to
	\be\label{particle action 2}
		S=\int \text{d}\tau\,\left(\la R\,g^{_{-1}}\dot{g}\ra +
		\la \tilde R\,\tilde g^{_{-1}}\dot{\tilde g}\ra- \frac{\xi}{2}
		\left(\la RR\ra+\la \tilde R \tilde R\ra+\m_{0}^2 \right)\right)~,
	\ee
	where $R$ and $\tilde R$ are Lie algebra valued phase space variables, $\xi$ becomes
	a Lagrange multiplier and its variation defines the mass-shell condition with timelike $R$
	\be\label{mass-shell}
		\la\, R\,R\, \ra +\la\, \tilde R\, \tilde R\, \ra +\m_{0}^2 =0~.
	\ee
	The Hamilton equations obtained from \eqref{particle action 2},
	\be\label{Hamilton eq}
		g^{_{-1}}\dot g=\xi R~,   \qquad \tilde g^{_{-1}}\dot{\tilde g}=\xi \tilde R~,  \qquad\quad
		\dot R=0~,        \qquad    \dot{\tilde R}=0~,
	\ee
	provide the conservation of $R$ and $\tilde R$, as well as of their `left' counterparts
	\be\label{L,L_s}
		L=g\,R\,g^{_{-1}}~, \quad  \qquad \tilde L=\tilde g\,\tilde R\,\tilde g^{_{-1}}~.
	\ee
	The dynamical integrals $L$, $\tilde L$, $R$ and $\tilde R$ are the Noether charges related to
	the invariance of the action \eqref{particle action 1} with respect to the isometry transformations \eqref{isometry tr}.

	The first order action  \eqref{particle action 2} defines the pre-symplectic form of the system
	\be\label{particle pre-symplectic form}
		\Theta=\la R g^{_{-1}} \mathrm{d}g\ra +
		\la \tilde R \tilde g^{_{-1}} \mathrm{d}\tilde g\ra~,
	\ee
	which leads to the following Poisson brackets 
	\be\ba\label{L-R PB}
		\qquad&\{L_a,\,L_b\}=2\e_{a b}\,^c \,L_c ~,&\qquad 
			&\{R_a  ,R_b \}=-2\epsilon_{a b}\,^c \,R_{c}~,&\qquad
			&\{L_a ,\,R_b\}=0~,&\\[.2em]
		&\{\tilde L_i,\,\tilde L_j\}=2\e_{ijk}\,\tilde L_k~,&
			&\{\tilde R_i,\,\tilde R_j\}=-2\e_{ijk}\,\tilde R_k~,&
			&\{\tilde L_i,\,\tilde R_j\}=0~,&
	\ea\ee
	where $L_a $, $\tilde L_j $, $R_a $, $\tilde R_j$ are the components of the charges in the bases \eqref{tt=} and \eqref{ss=}
	\be \label{components}
		L_a =\la\,\fkT_a \,L\,\ra~,\qquad \tilde L_j=\la\,{\tilde\fkT}_j\,\tilde L\,\ra~,\qquad
		R_a =\la\,\fkT_a \,R\,\ra~, \qquad \tilde R_j=\la\,{\tilde\fkT}_j\,\tilde R\,\ra~.
	\ee

	Since $R=R^a \fkT_a $ and $\tilde R=\tilde R_j\tilde\fkT_j$,
	the mass-shell condition \eqref{mass-shell} can be written as
	\be\label{constraint}
		R_a  R^a +\tilde R_j \tilde R_j+\m_{0}^2=0~,
	\ee
	and it obviously has vanishing Poisson brackets with components \eqref{components}.
	Hence, the components are gauge invariant and, therefore,
	the Poisson brackets algebra \eqref{L-R PB} will be preserved
	after gauge fixing.

	Let us choose the gauge $\xi=1$ and consider a solution of \eqref{Hamilton eq} in the ${\rm SU}(1,1)$ part
	\be\label{SU11 particle solution}
	g=e^{\m\tau\fkT_{0}}~, \qquad\qquad R=\m \fkT_{0}~,
	\ee
	which corresponds to the ${\rm AdS}_3$ particle of mass $\m \geq 0$ in the rest frame.
	The isometry transformations of \eqref{SU11 particle solution} provide a class of solutions
	parameterized by $\mu$ and the group variables
	\be\label{isometry map of solutions}
	g= g_{l}\,e^{\m\tau\fkT_{0}}\,g_{r}~, \qquad\qquad R=g_{r}^{_{-1}}\m \fkT_{0}\,g_{r}~.
	\ee
	To find the Poisson bracket structure on the space of
	parameters, we calculate the ${\rm SU}(1,1)$ part of the pre-symplectic form \eqref{particle pre-symplectic form}.
	For fixed $\tau$ this calculation yields
	\be\label{calculation of AdS 1-form}
	\th=\la R g^{_{-1}}\text{d} g\ra=\m\la\, {\fkT_{0}}\, g_{l}^{_{-1}}\,\text{d} g_{l}\,\ra+
	\m\la\, {\fkT_{0}}\, \text{d} g_{r}\,g_{r}^{_{-1}}\,\ra-\tau\m\text{d}\m~,
	\ee
	and we can neglect the exact form $-\tau\m\text{d}\m$. With help of the Cartan decomposition,
	\be\label{parametrization of g}
		g_{l}=e^{\a_{l}{\fkT_{0}}}\,e^{\g_{l}{\fkT_{1}}}\,e^{\b_{l}{\fkT_{0}}}~,\qquad\qquad
			g_{r}=e^{\b_{r}{\fkT_{0}}}\,e^{\g_{r}{\fkT_{1}}}\,e^{\a_{r}{\fkT_{0}}}~,
	\ee
	as elaborated in the appendix, the remaining terms in \eqref{calculation of AdS 1-form} reduces to a canonical 1-form
	\be\label{canonical AdS 1-form}
	\th=\m\text{d}\varphi+H_{l}\text{d}\phi_{l}+H_{r}\text{d}\phi_{r}~,
	\ee
	where we defined the following quantities, 
	\be\label{angle variables}
	\varphi=-(\a_{l}+\b_{l}+\a_{r}+\b_{r})~, \qquad \phi_{l}=\frac{\pi}{2}-2\a_{l}~, \qquad \phi_{r}={\pi}-2\a_{r}~,
	\ee
	\be\label{H_l,r}
		H_{l}=\frac{\m}{2}\,\big(\cosh (2\g_{l}) -1_{_{\,\!}}\big)~,\qquad\quad
			H_{r}=\frac{\m}{2}\,\big(\cosh (2\g_{r}) -1_{_{\,\!}}\big)~.
	\ee

	The conserved Noether charges constructed from \eqref{isometry map of solutions} and \eqref{parametrization of g} read
	\be\label{particle L and R}
		L=\mu\, e^{\a_{l}{\fkT_{0}}}\,e^{\g_{l}{\fkT_{1}}}\,{\fkT_{0}}\,
		e^{-\g_{l}{\fkT_{1}}}\,e^{-\a_{l}{\fkT_{0}}}~, \qquad
		R=\mu\, e^{-\a_{r}{\fkT_{0}}}\,e^{-\g_{r}{\fkT_{1}}}\,{\fkT_{0}}\,
		e^{\g_{r}{\fkT_{1}}}\,e^{\a_{r}{\fkT_{0}}}~,
	\ee
	and by \eqref{angle variables}-\eqref{H_l,r} their components become
	\be\ba\label{AdS dynamical integrals}
		&L^0=\m + 2H_{l}~,& \qquad     &R^0=\m + 2H_{r}~,&\\
		&L_\pm=\sqrt{\m H_{l}+H_{l}^2}\,e^{\pm i\phi_{l}}~,&\qquad &R_\pm=\sqrt{\m H_{r}+H_{r}^2}\,e^{\pm i\phi_{r}}~,&
	\ea\ee
	where $L_\pm=\frac{1}{2}(L_1\pm iL_2)$ and  $R_\pm=\frac{1}{2}(R_2\pm iR_1).$

	Similarly, for ${\rm SU}(2)$ we consider the isometry group orbit of the solution $\tilde g=e^{\tilde\m\tau\tilde\fkT_{3}}$, with $\tilde{\m}\geq0$. Repeating the same steps we obtain the canonical 1-form
	\be\label{canonical S 1-form}
		\tilde\th = \la \tilde R \tilde g^{_{-1}} \mathrm{d}\tilde g\ra=
		\tilde\m\text{d}\tilde\varphi+\tilde H_{l}\text{d}\tilde\phi_{l}+\tilde H_{r}\text{d}\tilde\phi_{r}~.
	\ee
	The canonical coordinates \eqref{S canonical coordinates}-\eqref{tilde H_l,r} given in the appendix parameterize the Noether charges $\tilde L_3$,
	$\tilde L_\pm=\frac{1}{2}(\tilde L_1\pm i\tilde L_2)$ and
	$\tilde R_3$, $\tilde R_\pm=\frac{1}{2}(\tilde R_2\pm i\tilde R_1)$ as follows
	\be\ba\label{S dynamical integrals}
	&\tilde L_3=\tilde\m-2\tilde H_{l}~,& \qquad &\tilde R _3=\tilde\m-2\tilde H_{r}~,&\\
	&\tilde L_\pm=\sqrt{\tilde\m \tilde H_l -\tilde H_{l}^2}\,\,e^{\pm i\tilde\phi_{l}}~,&\qquad &\tilde R_\pm=
	\sqrt{\tilde\m \tilde H_{r}-\tilde H_{r}^2}\,\,e^{\pm i\tilde\phi_{r}}~.&
	\ea\ee

	From the canonical variables $H \geq 0$ and $\phi \in {\rm S}^1$ one naturally defines creation-annihilation variables as
	\be\label{a,a^+}
	a^\dag=\sqrt{H}\,\,e^{i\phi}~, ~\qquad\qquad  a=\sqrt{H}\,\,e^{-i\phi}~.~~
	\ee

	The form of the functions \eqref{AdS dynamical integrals} and \eqref{S dynamical integrals} then dictates the realization of the isometry group generators in terms of creation-annihilation operators, which is known as the Holstein-Primakoff transformation \cite{Holstein:1940zp, Dzhordzhadze:1994np}. Thus, we have
	\begin{align}\nn
		\qquad\qquad&L^0=\m + 2a_{l}^\dag a_{l}~,& 
			&R^0=\m + 2a_{r}^\dag a_{r}~,&\\
		\label{AdS operators}
		&L_+=a_{l}^\dag\sqrt{\m +a_{l}^\dag a_{l}}~,&
			&R_+=a_{r}^\dag\sqrt{\m+a_{r}^\dag a_{r}}~,&\\
		\nn
		&L_-=\sqrt{\m +a_{l}^\dag a_{l}}\,\,a_{l}~,&
			&R_-=\sqrt{\m+a_{r}^\dag a_{r}}\,\,a_{r}~,&\\[.4 em]
		\nn
		&\tilde L_3=\tilde\m-2\tilde a^\dag_{l}\,\tilde a_{l}~,&
			&\tilde R _3 = \tilde\m-2\tilde a^\dag_{r}\,\tilde a_{r}~,&
			\\
		\label{S operators}
		&\tilde L_+=\tilde a_{l}^\dag\sqrt{\tilde\m -\tilde a^\dag_{l}\,\tilde a_{l}}~,&
			&\tilde R_+=\tilde a^\dag_{r}\sqrt{\tilde \m-
		\tilde a^\dag_{r}\,\tilde a_{r}}~,&\\
		\nn
		&\tilde L_-=\sqrt{\tilde\m -\tilde a^\dag_{l}\,\tilde a_{l}}\,\,\tilde a_{l}~,&
			&\tilde R_-=\sqrt{\tilde\m-\tilde a^\dag_{r}\,\tilde a_{r}}\,\,\tilde a_{r}~.&
	\end{align}

	These yield a representation of $\su_l(1,1)\oplus \su_r(1,1)\oplus \su_l(2)\oplus \su_r(2)$ with basis vectors
	\be
		\ket{\m,\tilde\m\,;\, k_l, k_r, \tilde{k}_l, \tilde{k}_r}
			= \ket{\m, k_{l}} \ket{\m, k_{r}} \ket{\tilde{\m}, \tilde{k}_{l}} \ket{\tilde{\m}, \tilde{k}_{r}}~,
	\ee
	where $k_{l, r}$, $\tilde\m$, and $\tilde{k}_{l,r}$ are non-negative integers and furthermore $\tilde{k}_{l,r} \leq \tilde{\m}$.

	The representation is characterized by the Casimir numbers
	\be\ba\label{Casimir}
		&C_{\rm AdS} = -L_a L^a = -R_a R^a = \m(\m-2)~,& \\
		&\quad~ C_{\rm S}=\tilde L_j \tilde L_j=\tilde R_j \tilde R_j=\tilde\m(\tilde\m+2)~,&
	\ea\ee
	which are related through the mass-shell condition \eqref{constraint}
	\be\label{mass-shell q}
		C_{\rm AdS}=C_{\rm S}+\m_{0}^2~,
	\ee
	and we find
	\be\label{minimal energy}
		\m 
			= 1 + \sqrt{\m_{0}^2 + (\tilde\m+1)^2}~.
	\ee

	Since translations along the ${\rm AdS}_3$ time direction correspond to rotations in the $(X^0, X^{0'})$ plane, the energy operator is given by
	\be\label{energy}
		E=\frac{1}{2}\left(L^0+R^0\right)~,
	\ee
	and from \eqref{AdS operators} we obtain the energy spectrum
	\be\label{energy spectrum}
		E=\m + k_{l} + k_{r}~.
	\ee
	Here, $\m$ is defined by \eqref{minimal energy} and corresponds to the lowest energy level for a given total angular momentum $\tilde\m$ on ${\rm S}^3$. Equations \eqref{minimal energy} and \eqref{energy spectrum} reproduces the result obtained in the covariant quantization or in the static gauge approach \cite{Dorn:2010wt}.

	In the following section we use a similar scheme to calculate the energy spectrum of ${\rm SU}(1,1)\times {\rm SU}(2)$ string solutions.

\subsection*{The Spinning String in ${\rm SU}(1,1)\times {\rm SU}(2)$}

	The Polyakov action for the ${\rm SU}(1,1)\times {\rm SU}(2)$ string is given by
	\be\label{Polyakov action}
		S=-\frac{\sqrt\l}{4\pi}\int  \text{d}\tau\,\text{d}\s\,\,
		\sqrt{-h}\,h^{\a\b}\,\Big(\la\,g^{_{-1}} \p_\a g\, g^{_{-1}} \p_\b g\,\ra +
		\la\,\tilde g^{_{-1}} \p_\a \tilde g\, \tilde g^{_{-1}} \p_\b \tilde g\,\ra\Big)~,
	\ee
	Here, $\l$ is a dimensionless coupling constant, which in context of the AdS/CFT correspondence playes the role of the 't Hooft coupling. In analogy to the case of the particle, for the closed string this action is equivalent to
	\begin{align} \label{Action}
		S=&\int \text{d}\tau \int_0^{2\pi} \frac{\text{d}\s}{2\pi}\,\bigg(\la \calR\,g^{_{-1}}\dot{g}\ra 
			+ \la \tilde \calR\, \tilde g^{_{-1}}\dot{\tilde g}\ra-\xi_2\Big(\la \calR\,g^{_{-1}}{g}'\ra 
			+ \la \tilde \calR\, \tilde g^{_{-1}}{\tilde g}'\ra\Big) \\[.2em] \nn
		&\qquad\qquad\quad
			-\frac{\xi_1}{2 \sqrt{\l}}\Big( \la\, \calR\,\calR\,\ra
			+ \la \tilde \calR\,\tilde \calR\ra 
			+ \l\la(g^{_{-1}}{g'})^2\ra
			+ \l\la(\tilde g^{_{-1}}{\tilde g'})^2\ra\Big)
	\bigg)~.
	\end{align}
	The Lagrange multipliers $\xi_1$ and $\xi_2$  are related to the worldsheet metric by
	\be\label{xi-1,2}
		\xi_1=-\frac{1}{\sqrt{-h}\,h^{\tau\tau}}~, \qquad\qquad \xi_2=-\frac{h^{\tau\s}}{h^{\tau\tau}},\qquad
	\ee
	and their variations provide the Virasoro constraints
	\be\ba\label{Virasoro}
		&\la\, \calR\,\calR\,\ra+\la \tilde \calR\,\tilde \calR\ra
			+ \l\la(g^{_{-1}}{g'})^2\ra
			+ \l\la({\tilde g}^{_{-1}}{{\tilde g}'})^2\ra=0~, \\[.2em]
		&\qquad\qquad\quad \la \calR\,g^{_{-1}}{g}'\ra
			+ \la \tilde \calR {\tilde g}^{_{-1}}\,{\tilde g}'\ra=0~.
	\ea\ee
	The conformal gauge corresponds to $\xi_1=1$ and $\xi_2=0$. In this case the equations of motion obtained from \eqref{Action} become
	\be\ba\label{EOM}
		\quad\qquad\qquad&\sqrt\l\,\,g^{_{-1}}\dot{g}=\calR~,&\qquad\qquad
			&\dot \calR=\sqrt\l(g^{_{-1}}{g'})'~,&\qquad\qquad\\[.2em]
		&\sqrt\l\,\,\tilde g^{_{-1}}\dot{\tilde g}=\tilde \calR~,&
			&\dot{\tilde \calR}= \sqrt\l({\tilde g}^{_{-1}}{\tilde g'})'~,&
	\ea\ee
	and they are equivalent to
	\be\label{Lagrange eq.}
	\p_\tau\left(g^{_{-1}}\dot{g}\right)=\p_\s\left(g^{_{-1}}{g}'\right)~,
	\qquad\quad \p_\tau\left({\tilde g}^{_{-1}}\dot{\tilde g}\right)=\p_\s\left({\tilde g}^{_{-1}}{\tilde g}'\right)~.
	\ee

	We now consider the following solution of these equations \cite{Jorjadze:2012rj}
	\begin{align}\label{AdS string solution}
		&g = \begin{pmatrix}
								\cosh\vartheta \,e^{i(e\tau+m\s)}~ & \sinh\vartheta \,e^{i(p\tau+n\s)}\\
								\sinh\vartheta \,e^{-i(p\tau+n\s)}& \cosh\vartheta \,e^{-i(e\tau+m\s)}
							\end{pmatrix}~,\\
		\label{S string solution}
		&\tilde g=\begin{pmatrix}
              ~\cos\tilde\vartheta \,e^{i(\tilde e\tau+\tilde m\s)} & i\sin\tilde\vartheta \,e^{i(\tilde p\tau+\tilde n\s)}\\
              i\sin\tilde\vartheta \,e^{-i(\tilde p\tau+\tilde n\s)}& \cos\tilde\vartheta \,e^{-i(\tilde e\tau+\tilde m\s)}
            \end{pmatrix}~,
	\end{align}
	with the parameters fulfilling
	\be\label{conditions for e,p}
		p^2-e^2=n^2-m^2~, \qquad\qquad \tilde p^2-\tilde e^2=\tilde n^2-\tilde m^2~,
	\ee
	which turns out to be the renowned spinning string solution \cite{Frolov:2003qc, Frolov:2003tu, Arutyunov:2003za}.

	In the appendix we present the matrices $g^{_{-1}}\dot{g}$, $g^{_{-1}}{g}'$, $\tilde g^{_{-1}}\dot{\tilde g}$, $\tilde g^{_{-1}}{\tilde g}'$ calculated from \eqref{AdS string solution}-\eqref{S string solution}. The corresponding equations \eqref{AdS R}-\eqref{S R} show that the conditions \eqref{conditions for e,p} indeed provide \eqref{Lagrange eq.}. The matrices $\calR$ and $\tilde \calR$ are defined by the Hamilton equations \eqref{EOM} and the Virasoro constraints \eqref{Virasoro} then lead to the additional conditions
	\be\ba\label{Virasoro 1}
		&(e^2+m^2)\cosh^2\vartheta-(p^2+n^2)\sinh^2\vartheta
			=(\tilde e^2 + \tilde m^2)\cos^2\tilde\vartheta
			+(\tilde p^2 + \tilde n^2)\sin^2\tilde\vartheta~,\\
		&\qquad\qquad\qquad 
			m e \cosh^2\th-np\sinh^2\vartheta
			= \tilde m \tilde e\cos^2\tilde\vartheta
			+ \tilde n\tilde p\sin^2\tilde\vartheta~,
	\ea\ee
	which are obtained from the induced metric \eqref{Tr dot-pime} given in the appendix.

	Note that the components of the induced metric tensor are constants on both the ${\rm SU}(1,1)$ and the ${\rm SU}(2)$ projections. The scheme of Pohlmeyer reduction \cite{Pohlmeyer:1975nb, Grigoriev:2007bu} for a flat induced metric yields a linear system with constant coefficients, which is simply integrated in the exponential form like in \eqref{AdS string solution}-\eqref{S string solution} \cite{Dorn:2010xt}. This is a typical feature of these so-called homogeneous solutions \cite{Frolov:2003qc, Frolov:2003tu, Arutyunov:2003za}.

	Since we consider a closed string in ${\rm SU}(1,1)\times {\rm SU}(2)$, the parameters $m$, $n$, $\tilde m$, $\tilde n$ have to be integer.
	However, if we unwrap the time coordinate, the polar angle in the $(X^0, X^{0'})$ plane, it has to be periodic in $\s$ itself. This is obviously achieved for $m=0$ only, which is assumed below.

	Thus, our solutions are parameterized by three winding numbers and six continuous variables, which satisfy the four conditions in \eqref{conditions for e,p}-\eqref{Virasoro 1}.
	Hence, for given winding numbers, we have a two parameter family of solutions.\footnote{Note that the particle solutions in ${\rm SU}(1,1)\times {\rm SU}(2)$ were parameterized by one variable $\tilde\m$.}

	Similarly to the particle dynamics, we consider the isometry group orbits of the solutions, with the aim to find their Hamiltonian description and quantization.
	For this purpose we analyze the pre-symplectic form defined by \eqref{Action},
	\be\label{string pre-symplectic}
		\Theta=\int_0^{2\pi} \frac{\text{d}\s}{2\pi}\,\Big(\la \calR\,g^{_{-1}}\text{d}{g}\ra +
		\la \tilde \calR\, \tilde g^{_{-1}}\text{d}{\tilde g}\ra\Big)~.
	\ee
	To calculate this 1-form on the space of orbits one has to make the  replacements
	\be\label{rplacements}
		\calR\, \mapsto \, \sqrt\l\,g_{r}^{_{-1}}\,g^{_{-1}}\dot g \,g_{r}~, \qquad g \, \mapsto \, g_{l}\,g\,g_{r}~,
		\qquad g^{_{-1}}\, \mapsto \, g_{r}^{_{-1}}g^{_{-1}}g_{l}^{_{-1}}~,
	\ee
	similarly for the ${\rm SU}(2)$ term, and then identify $g$ with \eqref{AdS string solution} and $\tilde g$ with \eqref{S string solution}, respectively. For the ${\rm SU}(1,1)$ part this yields
	\be\label{AdS strinng 1-form 1}
		\theta=\la\, L\, g_{l}^{_{-1}}\,\text{d} g_{l}\,\ra+
		\la\, R\, \text{d} g_{r}\,g_{r}^{_{-1}}\,\ra+
		\sqrt\l\int_0^{2\pi} \frac{\text{d}\s}{2\pi}\,\la \,g^{_{-1}}\dot g \,g^{_{-1}}\text{d}g\,\ra~,
	\ee
	where $L$ and $R$ are the Noether charges related to the isometries \eqref{isometry tr} as
	\be\label{AdS string charges}
		L=\sqrt\l\int_{0}^{2\pi}\frac{\text{d}\s}{2\pi}\,\,\dot{g}\,g^{_{-1}}~, \qquad\quad
			R=\sqrt\l\int_{0}^{2\pi}\frac{\text{d}\s}{2\pi}\,\,g^{_{-1}}\dot{g}~,
	\ee
	and the differential of $g$ in the last term of \eqref{AdS strinng 1-form 1} is taken with respect to the parameters of the solution \eqref{AdS string solution}.
	The calculations given by \eqref{1-form calculations} in the appendix show that the last term in \eqref{AdS strinng 1-form 1} is an exact form and can be neglected.
	The ${\rm SU}(2)$ part is computed in a similar way and alltogether we find the 1-form
	\be\label{final 1-form}
		\Theta=\la\, L\, g_{l}^{_{-1}}\,\text{d} g_{l}\,\ra
			+ \la\, R\, \text{d} g_{r}\,g_{r}^{_{-1}}\,\ra
			+ \la\, \tilde L\,\tilde g_{l}^{_{-1}}\,\text{d}\tilde g_{l}\,\ra
			+ \la\, \tilde R\, \text{d}\tilde g_{r}\,\tilde g_{r}^{_{-1}}\,\ra~,
	\ee
	where $\tilde L$ and $\tilde R$ are the Noether charges similar to \eqref{AdS string charges},
	\be\label{S string charges}
		\tilde L=\sqrt\l\int_{0}^{2\pi}\frac{\text{d}\s}{2\pi}\,\,\dot{\tilde g}\,\tilde g^{_{-1}}~, \qquad\quad
			\tilde R=\sqrt\l\int_{0}^{2\pi}\frac{\text{d}\s}{2\pi}\,\,\tilde g^{_{-1}}\dot{\tilde g}~.
	\ee
	These charges are easily calculable by the currents given in the appendix. However, their matrix form depends on the winding numbers and one has to distinguish between the cases $n\neq 0$ and $n=0$ for ${\rm SU}(1,1)$, as well as $\tilde m^2\neq\tilde n^2$ and $\tilde m^2=\tilde n^2$ for ${\rm SU}(2)$.

	Let us consider the case $n\neq 0$ and $\tilde m^2\neq\tilde n^2$.
	The integration of the off-diagonal terms of the currents \eqref{AdS R}-\eqref{dot L} vanish and we obtain
	\be\label{string charges 1}
		L=\m_{l}\,\fkT_{0}~, \qquad \tilde L=\tilde\m_{l}\,\tilde\fkT_{3}~, \qquad\quad
			R=\m_{r}\,\fkT_{0}~, \qquad \tilde R=\tilde\m_{r}\,\tilde\fkT_{3}~,
	\ee
	where
	\begin{align}\label{AdS left-right masses}
	\qquad&\m_{l}=\sqrt\l\,(e\cosh^2\vartheta-p\sinh^2\vartheta)~,&
		&\m_{r}=\sqrt\l\,(e\cosh^2\vartheta+p\sinh^2\vartheta)~,&\\
	\label{S left-right masses}
	&\tilde\m_{l}=\sqrt\l\,(\tilde e\cos^2\tilde\vartheta+\tilde p\sin^2\tilde\vartheta)~,&\qquad
		&\tilde\m_{r}=\sqrt\l\,(\tilde e\cos^2\tilde\vartheta-\tilde p\sin^2\tilde\vartheta)~.&
	\end{align}
	We can assume that the numbers $\m_{l}$, $\m_{r}$, $\tilde\m_{l}$, $\tilde\m_{r}$ are non-negative.

	Similarly to the particle case, the 1-form \eqref{final 1-form} then becomes
	\be\label{AdS strinng 1-form}
		\Theta=\m_{l}\la\, {\fkT_{0}}\, g_{l}^{_{-1}}\,\text{d} g_{l}\,\ra+
		\m_{r}\la\, {\fkT_{0}}\, \text{d} g_{r}\,g_{r}^{_{-1}}\,\ra+
		\tilde\m_{l}\la\, {\fkT_{3}}\,\tilde g_{l}^{_{-1}}\,\text{d}\tilde g_{l}\,\ra+
		\tilde\m_{r}\la\, {\fkT_{3}}\, \text{d}\tilde g_{r}\,\tilde g_{r}^{_{-1}}\,\ra~,
	\ee
	and the same parametrization as in \eqref{parametrization of g} leads to the canonical 1-form
	\be
		\Theta =
			\m_{l}\text{d}\varphi_{l}
			+ H_{l}\text{d}\phi_{l} + \m_{r}\text{d}\varphi_{l}
			+ H_{r}\text{d}\phi_{r} + \tilde\m_{l}\text{d}\tilde\varphi_{l}
			+ \tilde H_{l}\text{d}\tilde\phi_{l} 
			+ \tilde\m_{r}\text{d}\tilde\varphi_{r}
			+ \tilde H_{r}\text{d}\tilde\phi_{r}~.
	\ee
	The components of the symmetry generators have the same form as in \eqref{AdS dynamical integrals} and \eqref{S dynamical integrals}
	\begin{align}
		&\ba\label{AdS string dynamical integrals}
			&L^0=\m_{l} + 2H_{l}~,& \qquad     &R^0=\m_{r} + 2H_{r}~,&\\
			&L_\pm=\sqrt{\m_{l} H_{l}+H_{l}^2}\,e^{\pm i\phi_{l}}~,
			\,&\qquad &R_\pm=
			\sqrt{\m_{r} H_{r}+H_{r}^2}\,e^{\pm i\phi_{r}}~,&
		\ea\\[.2em]
		&\ba\label{S string dynamical integrals}
			&\tilde L_3=\tilde\m_{l}-2\tilde H_{l}~,& \qquad &\tilde R _3=\tilde\m_{r}-2\tilde H_{r}~,&\\
			&\tilde L_\pm=\sqrt{\tilde\m_{l} \tilde H_l -\tilde H_{l}^2}\,\,e^{\pm i\tilde\phi_{l}}~,&\qquad &\tilde R_\pm=
			\sqrt{\tilde\m_{r} \tilde H_{r}-\tilde H_{r}^2}\,\,e^{\pm i\tilde\phi_{r}}~,&
		\ea
	\end{align}
	Here, now the Casimir numbers $\m_{l}$ and $\m_{r}$ are independent, whereas $\tilde\m_{l}$ and $\tilde\m_{r}$ are integers of the same parity, which ensures that the total angular momentum $\frac{1}{2}(\tilde{\m}_l + \tilde{\m}_l)$ on ${\rm S}^3$ takes integer values.

	Hence, as in \eqref{AdS operators}-\eqref{S operators}, the Holstein-Primakoff transformation provides a realization of the isometry group generators and
	the energy given by \eqref{energy} is now obtained from \eqref{AdS string dynamical integrals} and  \eqref{AdS left-right masses}, having the spectrum
	\be\label{string spectrum}
		E=E_0 + k_{l} + k_{r}~,
	\ee
	where $k_{l}$ and $k_{r}$ are non-negative integers and $E_0=\sqrt\l\,e\cosh^2\vartheta$ corresponds to the minimal energy for given $\tilde\m_{l}$, $\tilde\m_{r}$. To find the dependence of this term on $\tilde\m_{l}$, $\tilde\m_{r}$, and the winding numbers one has to use \eqref{S left-right masses} and the constraints
	\eqref{conditions for e,p}-\eqref{Virasoro 1}.
	Hence, we get
	\be\label{e,p tilde}
		\tilde e=\f{1}{\sqrt\l}\,\f{\tilde\m_{l}+\tilde\m_{r}}{1+\cos2\tilde\vartheta}~, \qquad
		\tilde p=\f{1}{\sqrt\l}\,\f{\tilde\m_{l}-\tilde\m_{r}}{1-\cos2\tilde\vartheta}~.
	\ee

	Inserting them in \eqref{conditions for e,p}, one gets a fourth order equation for $\cos2\tilde\vartheta$
	\be\label{eq for cos th}
		(\tilde\m_{l}+\tilde\m_{r})^2(1-\cos2\tilde\vartheta)^2 - (\tilde\m_{l}-\tilde\m_{r})^2(1+\cos2\tilde\vartheta)^2 =
		\l(\tilde m^2-\tilde n^2)(1-\cos^22\tilde\vartheta)^2~.
	\ee
	The solution of this equation and \eqref{e,p tilde} define the right hand sides of \eqref{Virasoro 1} as a function of the coupling constant and four integers $(\tilde\m_{l},\tilde\m_{r},\tilde m,\tilde n)$. Solving \eqref{Virasoro 1} for $e^2$ and $\sinh\vartheta$ one obtains a third order equation, which can be solved in a standard way. Hence, we acquired exact quantization of the spinning string solution at hand, where however the final answer for $E_0$ takes a rather complicated form.

	Furthermore, in analogy to the discussion in \cite{Frolov:2013lva}, we expect that the obtained spectrum concurs with the one of corresponding states of the full superstring theory only at the leading order in strong coupling, $\l\gg1$. 
	Therefore, let us present the analysis only in this limit, which corresponds to the near-flat-space regime.

	First we consider the case when both $\tilde m$ and $\tilde n$ are non-zero and assume  $0<\tilde m^2<\tilde n^2$. Using \eqref{conditions for e,p} and \eqref{S left-right masses}, the system \eqref{Virasoro 1} can be written as
	\be\ba\label{Vir-1-1}
		&e^2-2n^2\sinh^2\vartheta=\tilde e^2+2\tilde n^2\sin^2\tilde\vartheta+\tilde m^2\cos2\tilde\vartheta~,\\
		&|n|\sqrt{e^2+n^2}\,\sinh^2\vartheta=|\tilde m(\tilde\m_l+\tilde\m_r)+\tilde n(\tilde\m_l-\tilde\m_r)|\l^{-{1}/{2}}~.
	\ea\ee

	At large $\l$, from \eqref{eq for cos th} and \eqref{e,p tilde} we find
	\be\label{large lambda}
		\cos2\tilde\vartheta=1-\f{|\tilde\m_{l}-\tilde\m_{r}|}{\sqrt{\tilde n^2-\tilde m^2}}\,\l^{-{1}/{2}} + \mathcal{O}(\l^{-1}) ~,
		\qquad \tilde e=\f{\tilde\m_{l}+\tilde\m_{r}}{2}\,\l^{-{1}/{2}}+ \mathcal{O}(\l^{-1})~,
	\ee
	and then \eqref{Vir-1-1} yields $\sinh^2\vartheta=\mathcal{O}( \l^{-{1}/{2}})$, $e=|\tilde m|+\mathcal{O}(\l^{-{1}/{2}})$ and  $E_0=|\tilde m|\,{\l}^{{1}/{2}}+\mathcal{O}(\l^0)$.

	The case $0<\tilde n^2<\tilde m^2$ is analyzed similarly. Its large $\lambda$ behavior is govern by
	\be\label{large lambda '}
		\cos2\tilde\vartheta=-1+\f{\tilde\m_{l}+\tilde\m_{r}}{\sqrt{ m^2-n^2}}\,\l^{-{1}/{2}}+ \mathcal{O}(\l^{-1}) ~,
		\qquad \tilde p=\f{\tilde\m_{l}-\tilde\m_{r}}{2}\,\l^{-{1}/{2}}+ \mathcal{O}(\l^{-1})~,
	\ee
	which again follows from \eqref{eq for cos th} and \eqref{e,p tilde}. Writing now the first equation of \eqref{Virasoro 1} as
	\be\label{Vir-1-2}
		e^2-2n^2\sinh^2\vartheta=\tilde p^2+2\tilde m^2\cos^2\tilde\vartheta-\tilde n^2\cos2\tilde\vartheta~,
	\ee
	we find $\sinh^2\vartheta=\mathcal{O}( \l^{-{1}/{2}})$, $e=|\tilde n|+\mathcal{O}(\l^{-{1}/{2}})$ and  $E_0=|\tilde n|\,{\l}^{{1}/{2}}+\mathcal{O}(\l^0)$.

	The analysis of the case $|\tilde m|=|\tilde n|$ is the most simple and it leads to the same answer. Thus, if $\tilde m\neq 0$ and $\tilde n\neq 0$, the leading order behavior of $E_0$ is given by
	\be\label{general E}
	E_0=\mbox{min}(|\tilde m|, |\tilde n|)\,\l^{{1}/{2}}+\mathcal{O}(\l^0)~.
	\ee

	Note that for $\tilde m=0=\tilde n$, from \eqref{Virasoro 1} one has $n=0$. The solution then becomes $\s$ independent and it describes the massless particle in ${\rm AdS}_3\times {\rm S}^3$.

	It remains to analyze the two cases $\tilde m=0$, $\tilde n\neq 0$ and $\tilde m\neq 0$, $\tilde n= 0$.
	In the first case the system \eqref{Virasoro 1} reduces to
	\be\label{eq for e}
		e^2-2n^2\sinh^2\vartheta=\tilde e^2+2\tilde n^2\sin^2\tilde\vartheta~, \quad
		\sqrt{n^2e^2+n^4}\sinh^2\vartheta=\sqrt{\tilde n^2\tilde e^2+\tilde n^4}\sin^2\tilde\vartheta~.
	\ee
	Here, one has to use the same large $\l$ behavior as in \eqref{large lambda}
	\be\label{large lambda1}
		\cos2\tilde\vartheta=1-\f{|\tilde\m_{l}-\tilde\m_{r}|}{|\tilde n|}\,\l^{-{1}/{2}}+ \mathcal{O}(\l^{-1}) ~,
		\qquad \tilde e=\f{\tilde\m_{l}+\tilde\m_{r}}{2}\,\l^{-{1}/{2}}+ \mathcal{O}(\l^{-1})~.
	\ee
	From \eqref{eq for e} we then find
	\bea\label{e=1}
		&\sinh^2\vartheta=\dfrac{|\tilde n(\tilde\m_{l}-\tilde\m_{r})|}{2n^2}\,\l^{-{1}/{2}}+ \mathcal{O}(\l^{-1})\,, \quad
		e^2={2|\tilde n(\tilde\m_{l}-\tilde\m_{r})|}\,\l^{-{1}/{2}}+\mathcal{O}(\l^{-1})\,,&\\
		\label{e=1b}
		&E_0=\sqrt{2|\tilde n(\tilde\m_{l}-\tilde\m_{r})|}\,\,\l^{{1}/{4}} + \mathcal{O}(\l^{-{1}/{4}})\,.&
	\eea
	Note that for $\tilde\m_{l}=\tilde\m_{r}$, the exact solution of the system takes the following simple form
	\be\label{m=m}
		\cos2\tilde\vartheta=1~, \qquad \tilde e={\tilde\m_{l}}\,\l^{{1}/{2}}=e~, \qquad \sinh^2\vartheta=0~,\qquad
		E_0=\tilde\m_{l}~,
	\ee
	and it corresponds to a particle solution in \eqref{AdS string solution}-\eqref{S string solution}.

	In the second case, $\tilde n=0$, the system \eqref{Virasoro 1} can be written in the form
	\be\label{eq 2 for e}
		e^2-2n^2\sinh^2\vartheta=\tilde p^2+2\tilde m^2\cos^2\tilde\vartheta~, \quad
		\sqrt{n^2e^2+n^4}\sinh^2\vartheta=\sqrt{\tilde m^2\tilde p^2+\tilde m^4}\cos^2\tilde\vartheta~.
	\ee
	The solutions of \eqref{eq for cos th} and \eqref{e,p tilde} at large $\l$ now are
	\be\label{large lambda 2}
		\cos2\tilde\vartheta=-1+\f{|\tilde\m_{l}+\tilde\m_{r}|}{|\tilde m|}\,\l^{-{1}/{2}}+ \mathcal{O}(\l^{-1}) ~,
		\qquad \tilde p=\f{\tilde\m_{l}-\tilde\m_{r}}{2}\,\l^{-{1}/{2}}+ \mathcal{O}(\l^{-1})~,
	\ee
	and \eqref{eq 2 for e} leads to
	\bea\label{e=2}
		&\sinh^2\vartheta=\dfrac{|\tilde m(\tilde\m_{l}+\tilde\m_{r})|}{2n^2}\,\l^{-{1}/{2}}+ \mathcal{O}(\l^{-1})\,, \quad
		e^2 = {2|\tilde m(\tilde\m_{l}+\tilde\m_{r})|}\l^{-{1}/{2}}+\mathcal{O}(\l^{-1})\,,\,&\\
		\label{e=2b}
		&E_0 = \sqrt{2|\tilde m(\tilde\m_{l}+\tilde\m_{r})|}\,\l^{{1}/{4}} + \mathcal{O}(\l^{-{1}/{4}})~.&
	\eea
	The case $\m_{l}=\m_{r}$ is again special, giving the simple solution in the $SU(2)$ part
	\be\label{m=m 2}
		\tilde p=0,  \quad\qquad \tilde e=|\tilde m|~, \quad\qquad \cos^2\tilde\vartheta=\f{\tilde\m_{l}}{|\tilde m|}\,\l^{-{1}/{2}}~,
	\ee
	and the corresponding minimal energy
	\be\label{min E}
	E_0=2\sqrt{|\tilde m \tilde\m_{l}|}\,\l^{{1}/{4}}\,+\mathcal{O}(\l^{-{1}/{4}})~.
	\ee

	Note that \eqref{eq for e}-\eqref{e=1b} become \eqref{eq 2 for e}-\eqref{e=2b} by substituting
	\be
		\{\tilde{n}, \tilde{e}, \tilde{\vartheta}, \tilde{\m}_l + \tilde{\m}_l \}
			\quad\longleftrightarrow\quad
		\{\tilde{m}, \tilde{p}, \tilde{\vartheta} + \tfrac{\pi}{2}, \tilde{\m}_l - \tilde{\m}_l \}~,
	\ee
	which corresponds to interchanging the $(Y^1, Y^2)$ with the $(Y^3, Y^4)$ plane.\footnote{This symmetry could have been made manifest by also allowing for negative $\su(2)$ Casimir numbers, $\tilde{\m}_{l,r} \in \mathbb{Z}$ with $\frac{1}{2}(\tilde{\m}_l + \tilde{\m}_r) \in \mathbb{Z}$, which however would have complicate most formulas.}

	We can now compare the results qualitatively. For $m, n\!>\!0$ the scaling of the minimal energy \eqref{general E}, $E_0 \propto \l^{1/2}$, suggests that the corresponding strings are long. In contrast, for $m=0$ or $n=0$ we found $E_0 \propto \l^{1/4}$, see \eqref{e=1b} and \eqref{e=2b}, which is the typical scaling behavior of short strings. Indeed, in the first case the string wraps both circles, the one in the $(Y^1, Y^2)$ and the one in the $(Y^3, Y^4)$ plane, and hence cannot become small, while for the latter cases this is possible, as the string now only wraps one circle.

	As we are particularly interested in the short string regime, recall that the $\su(2)$ Casimir numbers $\tilde{\m}_l$ and $\tilde{\m}_r$ have the same parity. Hence, \eqref{e=1b} and \eqref{e=2b} both yield the minimal energy of the form
	\be
		E_0 = 2 \sqrt{N} \l^{-1/4} + \mathcal{O}(\l^{1/4})~,\qquad\text{with}\quad N\in \mathbb{N}~.
	\ee 
	which is nothing but the renowned result by Gubser, Klebanov, and Polyakov \cite{Gubser:2002tv} for the near-flat-space limit. The lowest excited states, $N=1$, ought to be dual to some members of the Konishi multiplet.


$\,$\\
{\bf \large Acknowledgments}

\vspace{3mm}

\noindent
	We thank Harald Dorn, Sergei Frolov, Ben Hoare and Jan Plefka for useful discussions, and Stijn van Tongeren for valuable comments on the manuscript. We also thank Chrysostomos Kalousios and Zurab Kepuladze for collaboration at an initial stage of the work. G.J. thanks the Humboldt University of Berlin and the Max-Planck Institute for Gravitational Physics in Potsdam for kind hospitality. M.H. thanks Nordita in Stockholm for kind hospitality. The research leading to these results has received funding from the Volkswagen-Foundation, WFS, Rustaveli GNSF, the International Max Planck Research School for Geometric Analysis, Gravitation and String Theory, and a DFG grant in the framework of the SFB 647.

\newpage

\subsection*{Appendix}

	The commutation relations of the basis vectors \eqref{su(1,1) basis} provide the following adjoint transformation properties
	\be\label{tr of b}
		e^{\g\fkT_{1}}\,\fkT_{0}\,e^{-\g\fkT_{1}}=\cosh(2\g)\,\fkT_{0}+\sinh(2\g)\,\fkT_{2}~,
	\quad~
		e^{\a\fkT_{0}}\,\fkT_{2}\,e^{-\a\fkT_{0}}=
		\cos(2\a)\,\fkT_{2}+\sin(2\a)\,\fkT_{1}~.
	\ee
	The conserved charges \eqref{particle L and R} then can be written as
	\be\ba\label{particle L and R=}
		&L= \m\,\Big(\!\cosh(2\g_{l})\,\fkT_{0}+\sinh(2\g_{l}) \big(\cos(2\a_{l})\,\fkT_{2} + \sin(2\a_{l})\,\fkT_{1}\big)_{_{\!}}\Big)~,\\
		&R= \m\,\Big(\!\cosh(2\g_{r})\,\fkT_{0}-\sinh(2\g_{l}) \big(\cos(2\a_{r})\,\fkT_{2} - \sin(2\a_{r})\,\fkT_{1}\big)_{_{\!}}\Big)~.
	\ea\ee
	From these equations follow that the angle variables $\phi_{l}$ and $\phi_{r}$ defined in \eqref{angle variables} correspond to the phases of $L_1+iL_2$ and $R_2+iR_1$, respectively, as in \eqref{AdS dynamical integrals}.

	By \eqref{parametrization of g}, the `left' term of the 1-form \eqref{calculation of AdS 1-form} becomes
	\be\label{L-1-form}
		\m\la\, \fkT_{0}\, g_{l}^{_{-1}}\,\text{d} g_{l}\,\ra=
		\m\Big(\la\, e^{\g_{l}\fkT_{1}}\,\fkT_{0}\,
		e^{-\g_{l}\fkT_{1}}\, \fkT_{0}\,\ra\,\text{d}\a_{l}
		- \text{d}\b_{l}
		\Big)~.
	\ee
	with the coefficient of $\text{d} \g_{l}$ being $\la\,\fkT_{0}\,\fkT_{1}\,\ra=0$. Similarly, the 'right' term in \eqref{calculation of AdS 1-form} reads
	\be\label{R-1-form}
		\m\la\, \fkT_{0}\,\text{d} g_{r}\, g_{r}^{_{-1}}\,\ra=
		\m\Big(\la\, e^{\g_{r}\fkT_{1}}\,\fkT_{0}\,
		e^{-\g_{r}\fkT_{1}}\, \fkT_{0}\,\ra\,\text{d}\a_{r}
		- \text{d}\b_{r}
		\Big)~.
	\ee
	Taking into account then \eqref{tr of b} and $\la\,\fkT_a\,\fkT_b\,\ra = \eta_{a b}$, we arrive at \eqref{canonical AdS 1-form}-\eqref{H_l,r} and \eqref{AdS dynamical integrals}. Note that in \eqref{H_l,r} we substracted 1 from $\cosh(2\g)$ to have $H \geq 0$.

	For ${\rm SU}(2)$ one has transformations similar to \eqref{tr of b},
	\be\label{tr of b tilde}
		e^{\tilde\g\tilde\fkT_{1}}\,\tilde\fkT_{3}\,e^{-\tilde\g\tilde\fkT_{1}}
			= \cos(2\tilde\g)\,\tilde\fkT_{3} + \sin(2\tilde\g)\,\tilde\fkT_{2}~,
		\qquad
		e^{\tilde\a\tilde\fkT_{3}}\ \tilde\fkT_{2}\,e^{-\tilde\a\tilde\fkT_{0}}
		= \cos(2\tilde\a)\,\tilde\fkT_{2}+\sin(2\tilde\a)\,\tilde\fkT_{1}~.
	\ee

	Repeating the same steps as for ${\rm SU}(1,1)$, one obtains equations similar to \eqref{particle L and R=}-\eqref{R-1-form}, where one has to substitute untilded by tilded parameters along with the replacements
	\be\label{replacement rule}
		\cosh(2\g) \mapsto \cos(2\tilde\g) ~, \qquad 
			\sinh(2\g) \mapsto \sin2\tilde\g ~, \qquad 
			\fkT_{0}\mapsto \tilde\fkT_{3}~.
	\ee
	This procedure yields the following 1-form
	\be\label{su(2) 1-form }
		\tilde\theta
			= \tilde\m(\text{d}\tilde\b_{l}+\text{d}\tilde\b_{r})
			+ \tilde\m \cos(2\tilde\g_{l})\,\text{d}\tilde\a_{l}
			+ \tilde\m \cos(2\tilde\g_{r})\,\text{d}\tilde\a_{r}~,
	\ee
	which takes the canonical form \eqref{canonical S 1-form} with
	\bea\label{S canonical coordinates}
		&\tilde\varphi 
			= \tilde\a_{l}+\tilde\b_{l}+\tilde\a_{r}+\tilde\b_{r}~, \qquad
		\tilde\phi_{l}=\dfrac{\pi}{2}-2\tilde\a_{l}~,
		\qquad \tilde\phi_{r}={\pi}-2\tilde\a_{r}~,&\\[.2em]
		\label{tilde H_l,r}
		&\tilde H_{l}=\dfrac{\tilde \m}{2}\,[1-\cos (2\tilde\g_{l})]~,\qquad\quad  \tilde H_{r}=\dfrac{\tilde \m}{2}\,[1-\cos (2\tilde\g_{r})]~.&
	\eea
	The ${\rm SU}(2)$ version of \eqref{particle L and R=} then provides \eqref{S canonical coordinates}.

	To check that \eqref{AdS string solution} and \eqref{S string solution}, together with \eqref{conditions for e,p}, satisfy equations \eqref{Lagrange eq.}, we calculate the left-invariant currents and find
	\begin{align}
		\label{AdS R} 
		&\ba
		&g^{_{-1}}\dot{g}=\frac{i}{2}
			\begin{pmatrix}
				(e-p) + (e+p)\cosh2\vartheta  
				& (e+p)\,e^{-i\,\om_-} \sinh 2\vartheta \\
				-(e+p)\,e^{i\,\om_-} \sinh 2\vartheta 
				& -(e-p) - (e+p)\cosh2\vartheta
			\end{pmatrix}, \\[.2em]  
		&g^{_{-1}}{g}'=\frac{i}{2}
			\begin{pmatrix}
				(m-n) + (m+n)\cosh2\vartheta 
				& (m+n)\,e^{-i\,\om_-} \sinh 2\vartheta \\
				-(m+n)\,e^{i\,\om_-} \sinh 2\vartheta
				& -(m-n)-(m+n)\cosh2\vartheta
			\end{pmatrix},
		\ea \\[.6em]
		\label{S R}
		&\ba
		&\tilde g^{_{-1}}\dot{\tilde g}=\frac{i}{2}
			\begin{pmatrix}
				(\tilde e-\tilde p) + (\tilde e+\tilde p)\cos2\tilde\vartheta
					&i(\tilde e+\tilde p)\sin 2\tilde\vartheta \,e^{-i\,\tilde{\om}_-}\\
				-i(\tilde e+\tilde p)\sin 2\tilde\vartheta\,e^{i\,\tilde{\om}_-}
					&-(\tilde e-\tilde p)-(\tilde p+\tilde e)\cos2\tilde\vartheta
			\end{pmatrix}, \\[.2em] 
		&\tilde g^{_{-1}}{\tilde g}'=\frac{i}{2}
			\begin{pmatrix}
				(\tilde m-\tilde n) + (\tilde m+\tilde n)\cos2\tilde\vartheta&
					i(\tilde m+\tilde n)\,e^{-i\,\tilde{\om}_-}\sin 2\tilde\vartheta\\
				-i(\tilde m+\tilde n)\,e^{ i\,\tilde{\om}_-}\sin 2\tilde\vartheta&
					-(\tilde m-\tilde n) - (\tilde m+\tilde n)\cos2\tilde\vartheta
			\end{pmatrix},
		\ea
	\end{align}
	with the abbreviations $\om_\pm =(e \pm p)\tau + (m \pm n)\s$ and  $\tilde{\om}_\pm =(\tilde e\pm\tilde p)\tau+(\tilde m\pm\tilde n)\s$.

	Since the diagonal components of these matrices are constants, one has to check \eqref{Lagrange eq.} for the off-diagonal entries only, giving the conditions \eqref{conditions for e,p}.

	Similar calculations for the right-invariant currents 
	yield
	\be\label{dot L}
		\ba
		&\dot{g}\,g^{_{-1}}=\frac{i}{2}
			\begin{pmatrix}
				(e+p) + (e-p)\cosh2\vartheta
				&-(e-p)\,e^{i\,\om_+} \sinh 2\vartheta\\
				(e-p)\,e^{-i\,\om_+} \sinh 2\vartheta
				&-(e+p) -(e-p)\cosh2\vartheta
            \end{pmatrix},\quad~~\\[.2em]
		&\dot{\tilde g}\,\tilde g^{_{-1}}=\frac{i}{2}
			\begin{pmatrix}
				(\tilde e+\tilde p) + (\tilde e-\tilde p)\cos2\tilde\vartheta
				&-i(\tilde e-\tilde p)\,e^{i\,\tilde{\om}_+} \sin 2\tilde\vartheta\\
				i(\tilde e-\tilde p)\,e^{-i\,\tilde{\om}_+}\sin 2\tilde\vartheta
				&-(\tilde e+\tilde p) - (\tilde e-\tilde p)\cos2\tilde\vartheta
			\end{pmatrix}.
	\ea\ee

	The induced metric tensor components obtained from \eqref{AdS R}-\eqref{S R} read
	\begin{align}\nn
		&\la\, (g^{_{-1}}\dot{g})^2\,\ra = p^2\sinh^2\vartheta-e^2\cosh^2\vartheta~,&
		&\la\, (\tilde g^{_{-1}}\dot{\tilde g})^2\,\ra = \tilde p^2\sin^2\tilde\vartheta + \tilde e^2\cos^2\tilde\vartheta~,& \\
		\label{Tr dot-pime}
		&\la\, (g^{_{-1}}{g}')^2\,\ra = n^2\sinh^2\vartheta-m^2\cosh^2\vartheta~,& 
		&\la\, (\tilde g^{_{-1}}{\tilde g}')^2\,\ra = \tilde n^2\sin^2\tilde\vartheta+\tilde m^2\cos^2\tilde\vartheta~,& \\ 
		\nn
		&\la\, g^{_{-1}}\dot{g} \,g^{_{-1}}{g}')\,\ra=n p\sinh^2\vartheta-m e\cosh^2\vartheta~,&
		&\la\,\tilde g^{_{-1}}\dot{\tilde g}\, \tilde g^{_{-1}}{\tilde g}'\,\ra=\tilde n\tilde p\sin^2\tilde\vartheta+\tilde
		m\tilde e\cos^2\tilde\vartheta~.&
	\end{align}
The Virasoro constraints \eqref{Virasoro} are expressed through these components and one gets additional conditions
on the parameters given by \eqref{Virasoro 1}.

Finally, we present formulae useful for calculations of the pre-symplectic form
\be\ba
&\la\, g^{_{-1}}\dot{g} \,\,g^{_{-1}}\partial_\vartheta{g})\,\ra=0~,&
\qquad
&\la\, g^{_{-1}}\dot{g} \,\left(p\,g^{_{-1}}\partial_e{g}+e\,g^{_{-1}}\partial_p{g})\right)\,\ra=-ep\tau~,&
\\
\label{1-form calculations}
&\la\,\tilde g^{_{-1}}\dot{\tilde g} \,\,\tilde g^{_{-1}}\partial_{\tilde\vartheta}{\tilde g})\,\ra=0~, &
\qquad
&\la\, \tilde g^{_{-1}}\dot{\tilde g} \,\left(\tilde p\,\tilde g^{_{-1}}\partial_{\tilde e}{\tilde g}+
\tilde e\,\tilde g^{_{-1}}\partial_{\tilde p}{\tilde g})\right)\,\ra=\tilde e\tilde p\tau~.
\ea\ee
Here $g$ and $\tilde g$ are given again by \eqref{AdS string solution}-\eqref{S string solution} and the
calculation is straightforward.
Taking into account then the constraint \eqref{conditions for e,p} between the parameters $e$ and $p$,
the last term in \eqref{AdS strinng 1-form 1} becomes  the exact form -$\tau e\text{d}e$.
Obviously, the same is valid for the ${\rm SU}(2)$ part.
\newpage

\bibliographystyle{nb}
\bibliography{bibSpires}

\end{document}